\documentclass[journal]{IEEEtran} 
                           
\bibliography{IEEEabrv,mybib}

\usepackage{amsmath}
\usepackage{amssymb}
\usepackage{amsthm}
\usepackage{xfrac}
\usepackage{graphicx}
\usepackage{epstopdf}
\usepackage{textcomp}
\usepackage{multicol}
\usepackage{ulem}
\usepackage{color}
\usepackage{tabularx}
\usepackage[noadjust]{cite}
\usepackage{algorithm}
\usepackage{algorithmicx}
\usepackage[noend]{algpseudocode}

\newtheorem{propapp}{Proposition}
\newtheorem{definition}{Definition}

\title{Graph-Variate Signal Analysis}
\author{
	Keith Smith$^{1,2,3,*}$, Loukianos Spyrou$^{1}$, \textit{Member, IEEE}, \& Javier Escudero$^{1}$, \textit{Member, IEEE}
	\thanks{$^{1}$School of Engineering, Institute for Digital Communications, University of Edinburgh, Alexander Graham Bell Building, Edinburgh, EH9 3FG, UK}
	\thanks{$^{2}$Alzheimer Scotland Dementia Research Centre, Psychology Department, University of Edinburgh, 7 George Square, Edinburgh, EH8 9JZ, UK}
	\thanks{$^{3}$Centre for Medical Informatics, Usher Institute, University of Edinburgh, 9 BioQuarter, Edinburgh, EH16 4UX, UK}
	\thanks{$^{*}$KS was supported by an Engineering and Physical Sciences Research Council (EPSRC) studentship and by Health Data Research UK (MRC ref Mr/S004122/1), which is funded by the UK Medical Research Council, EPSRC, Economic and Social Research Council, National Institute for Health Research (England), Chief Scientist Office of the Scottish Government Health and Social Care Directorates, Health and Social Care Research and Development Division (Welsh Government), Public Health Agency (Northern Ireland), British Heart Foundation and Wellcome. E-mail: k.smith@ed.ac.uk}%
}
	
\begin{document}
	
\maketitle%

\begin{abstract}
Incorporating graphs in the analysis of multivariate signals is becoming a standard way to understand the interdependency of activity recorded at different sites. The new research frontier in this direction includes the important problem of how to assess dynamic changes of signal activity. We address this problem in a novel way by defining the \textit{graph-variate signal} alongside methods for its analysis. Essentially, graph-variate signal analysis leverages graphs of reliable connectivity information to filter instantaneous bivariate functions of the multivariate signal. This opens up a new and robust approach to analyse joint signal and network dynamics at sample resolution. Furthermore, our method can be formulated as instantaneous networks on which standard network analysis can be implemented. When graph connectivity is estimated from the multivariate signal itself, the appropriate consideration of instantaneous graph signal functions allows for a novel dynamic connectivity measure-- \textit{graph-variate dynamic (GVD) connectivity}-- which is robust to spurious short-term dependencies. Particularly, we present appropriate functions for three pertinent connectivity metrics-- correlation, coherence and the phase-lag index. We show that our approach can determine signals with a single correlated couple against wholly uncorrelated data of up to 128 nodes in signal size (1 out of 8128 weighted edges). GVD connectivity is also shown to be more robust than i) other GSP approaches at detecting a randomly traveling spheroid on a 3D grid and ii) standard dynamic connectivity in determining differences in EEG resting-state and task-related activity. We also demonstrate its use in revealing hidden depth correlations from geophysical gamma ray data. We expect that the methods and framework presented will provide new approaches to data analysis in a variety of applied settings.
\end{abstract}
	
\section{Introduction}
Network science provides a well tried and tested framework for analysing graph topologies derived from pairwise dependencies between the agents, recordings or information received at different points of a given space \cite{Newm2010,Bara2016}. An interesting case arises when graphs are known or otherwise constructed for use in the analysis of multivariate signals, where each signal is associated with a node of the graph. Notably, the recently developed theory of Graph Signal Processing (GSP) outlines a promising approach to tackle such scenarios \cite{Shum2013, Sand2013}. In this setting, a signal, whose samples occur at graph nodes, is processed over the graph topology.

GSP is mainly concerned with the development of a cohesive signal processing theory for graph signals, analogous to classical signal processing \cite{Shum2013}. Spectral graph techniques are implemented, using the Eigen-decomposition of either the graph adjacency matrix \cite{Sand2013} or its Laplacian \cite{Shum2013}. These are then used to process graph signals through the Graph Fourier Transform (GFT) which is propositioned as analogous to the classical Fourier transform in standard signal processing. This approach has been applied in topics such as big data \cite{Sand2014} and neuroscience \cite{Riu2016}. Recent work on the integration of the temporal domain within the GSP framework is also underway \cite{Louk2016, Grassi2017}. This spectral approach, however, presents hurdles in interpretation in light of the fact that the frequencies of the graph signal emerge through graph eigenvectors which relate to a still unquantified extent to the graph topology. Further, the graph signal itself remains a passive component in the analysis treated as a vector separate from the graph adjacency matrix.
	
On the other hand, the Dirichlet energy of a graph signal \cite{Shum2013}, defined as
\begin{equation}\label{DE}
\mathbf{x}^{T}\mathbf{Lx} = \sum_{i,j = 1}^{n}w_{ij}(x_{i}-x_{j})^{2}
\end{equation}
for graph weights $w_{ij}$ and graph signal $\mathbf{x}$, where $\mathbf{L}$ is the graph Laplacian, is a more directly extracted feature of signal variability over the graph. In a recent reconceptualisation, it has shown promise as a way to measure dynamic connectivity of brain function from EEG recordings, where a graph of pairwise signal correlations was considered as a support for graph signals of instantaneous EEG activity \cite{Smit2016}. This described how the relationship between the Pearson correlation coefficient and the squared difference of the graph signals, i.e. the two components of \eqref{DE} in this instance (taking Pearson's correlation coefficient as the graph weight $w_{ij}$), complemented each other to provide a high temporal resolution connectivity measure.

Although this begins to reconceptualise how the Dirichlet energy can be treated, work is required to i) generalise this method for more pertinent connectivity measures available, ii) understand the further scope of possibilities enabled for analysis, and iii) understand the links and possible overlap between this line of enquiry and the established framework of GSP. With respect to points (i) and (ii) we expound upon this new conceptualisation of Dirichlet energy to establish a general methodology for studying bivariate functions of graph signal activity-- graph-variate signal analysis. This includes generalising dynamic connectivity estimation for various different connectivity measures and-- noting that graph-variate signal analysis can be framed in terms of adjacency matrices-- posing a form of network analysis conducted at sample resolution of the multivariate signal. With respect to point (iii), in the Appendix we layout a general mathematical framework of multivariate signals and graphs.  Following from this, we demonstrate that the classical GSP framework, where one deals with graph adjacency matrices (or the Laplacian) and graph signal vectors and their matrix multiplications, is indeed not adequate to consider general bivariate functions. This helps to realise graph-variate signal analysis as a new branch of analysis with distinct motivations and aims.

The newly proposed dynamic connectivity estimation is timely in light of current efforts in estimating dynamic connectivity from multivariate signals. A large contingent of research solutions for temporal networks take the form of events occurring at edges (i.e., between two nodes) which change over time. This is geared towards data in which node-specific activity is either not available or not meaningful \cite{Holme2012}. Such outputs are also well suited to the analysis of multi-layer networks, where each layer is composed of a network of connectivity at time $t$ and layers are arranged in a tensor \cite{Dedom2013}. For a multivariate signal, on the other hand, each of its univariate signals is directly associated with a node. Indeed, the graph itself is often constructed from pairwise dependencies between the signals. Nonetheless, attempts have been made to devise temporal and multi-layer network methods to analyse multivariate signals.
	
\begin{figure}[!tb]
	\centering
	\includegraphics[trim = 0 0 0 0,clip,scale=0.37]{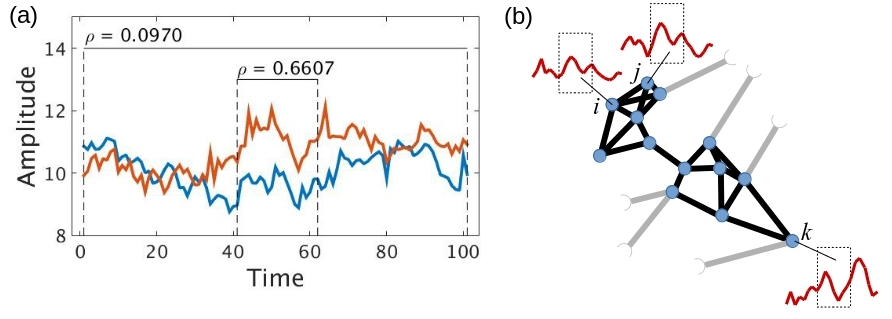}
	\caption{(a) A specific example of spurious short-term correlation coefficient, $\rho$, from independent realisations of an autoregressive model. (b) An illustration of how long-term connectivity weighting (dark edges connecting nodes) improves robustness of analysis of short-term transient dynamics. Nodes $i$, $j$ and $k$ all exhibit similar behaviour in the windowed epoch. However, from the topology of long-term connectivity it is clear that the correlation between $i$ and $j$, with a shortest path of 1, is more meaningful whereas correlations between $i$/$j$ and $k$, with shortest paths of 6, are likely spurious and should be disregarded.
	}
	\label{illustration}
\end{figure}

Neuroimaging is a prime example of this. Here, activity is often recorded at sensors (MEG/EEG) or voxels (fMRI) and topological dependencies are estimated via time-series correlations or phase dependencies. Suitable methods for the temporal analysis of networks in neuroscience is noted as an important open topic to gain a foothold on changing connectivity patterns \cite{FalRev, PapRev}. Most recent studies go the route of implementing disjoint \cite{Doron2012} or overlapping \cite{Leonardi2015, Braun2015, Braun2016} windows to construct a number of distinct chronologically separated graphs. This approach, however, is limited by the length of the window-- the less samples used to define the network, the less reliable is the connectivity estimate. Fig. \ref{illustration}(a) illustrates this, showing independent realisations of an autoregressive process in which spurious strong correlations (computed using Pearson's correlation coefficient $\rho$) can be found in short windows. On the other hand, the larger the window used the less meaningful it is at determining temporally refined connectivity estimates. Therefore obtaining reliable transient information is difficult.

Though one may consider instantaneous phases alone \cite{Glerean2012} which do allow signal resolution analysis, they are wide open to spurious connections, Fig. \ref{illustration} (a), and noisy fluctuations in the signal. In order to ameliorate this, we propose an approach via graph-variate signal analysis named graph-variate dynamic (GVD) connectivity, which maintains sample resolution. Essentially, our approach seeks to negate spurious short-term effects by weighting analysis with stable graph dependencies. This emphasises transient signal dynamics at the strongest connections and suppresses those from weak connections. An illustration of this is shown in Fig \ref{illustration}(b), where transient activity common to $i$ and $j$ should be regarded with more credence than if it were between $k$ and either $i$ or $j$. By choosing appropriate signal functions in the graph-variate signal analysis, we provide reasonable dynamic connectivity estimates for amplitude, power and phase-based connectivity in the form of the correlation coefficient, coherence and phase-lag index, respectively.

We demonstrate our methodology by determining its ability to correctly identify the presence of correlations in large datasets. This is done for various sizes of multivariate signals generated from an autoregressive process from which only a single truly coupled source exists. We then show that our approach outperforms state-of-the-art dynamic connectivity methods in an EEG resting state paradigm and show how our methods can be used as an investigative tool in geophysical exploration. Furthermore, outside of GVD connectivity, i.e. where the graph is constructed separately from the multivariate signal, we demonstrate how the more refined analyses enabled by our generalisation provides greater accuracy than comparable GSP approaches in a simple randomly travelling spheroid detection problem.

Our main aims are to introduce the general theoretical setting for graph-variate signal analysis (sections II \& III) and to provide evidence for the benefits of the applications of this theory over comparable benchmark approaches in simulations and applications (section IV). Our methods are geared to jointly answer what the stable connections in the data are and how the data behave instantaneously. These are, until now, usually sought separately. Therefore, we focus on exploring the possibilities encompassed by graph-variate signal analysis, comparing with relevant benchmarks in basic setups.

\section{Methods}
\subsection{Graph-variate signal analysis}
For reference, a table of the notation used for graph-variate signal analysis in this article is provided in Table \ref{NotTable}. Let $G = (\mathcal{V},\mathcal{E},\mathbf{W})$ be a graph with node set $\mathcal{V}= \{1,2,\dots, n\}$, edge set $\mathcal{E} = \{(i,j): i,j\in\mathcal{V}\}$, and corresponding weighted adjacency matrix $\mathbf{W}$ with entries $w_{ij}$ the weight of edge $(i,j)$. Also, let $\mathbf{X}\in\mathbb{R}^{n\times p}$ be a multivariate signal of size $n$ and length $p$. Firstly, we shall define a new mathematical object to denote a graph-variate signal. 
\begin{definition}
$\Gamma = (\mathcal{V}, \mathbf{X}, \mathcal{E}, \mathbf{W}) = (\mathbf{X},G)$ is a graph-variate signal where $\mathcal{V}$ is the set of nodes with $|\mathcal{V}|=n$; $\mathbf{X}\in 	\mathbb{R}^{n\times p}$ the multivariate signal indexed by $\mathcal{V}$; $\mathcal{E}$ the set of edges with $|\mathcal{E}|=2m$; and $\mathbf{W} = \{w_{ij}\}_{(i,j)\in \mathcal{E}}\in \mathbb{R}^{n\times n}$ the weighted adjacency matrix encoding a relevant topology in which the multivariate signal is set.
\end{definition}
At first this object may seem trivial as it concerns a grouping of things which are known. However, there is an important conceptual distinction from GSP here in that the graph signal is integrated into the object rather than defined separately from the graph. In this way, the object unifies a multivariate signal and graph as one object to be studied, rather than as two objects-- the graph and the graph signal-- interacting with one another.

\begin{table}[!tb]
	\caption{Notation for graph-variate signal analysis}
	\label{NotTable}
	\centering
\begin{tabular}{|c|l||c|l|}
	\hline
	$G$ & Graph & $F_{\mathcal{V}}$ & Node space function\\
	\hline
	$\mathbf{W}$ & Weighted adjacency matrix & $F_{\mathcal{E}}$ & Edge space function\\
	\hline
	$w_{ij}$ & $ij$th entry of $\mathbf{W}$ & $\underline{\mathbf{J}}$ & Node function tensor\\
	\hline
	$\mathbf{X}$ & Multivariate signal & $\underline{\mathbf{\Delta}}$ & Graph-variate network\\
	\hline
	$\mathbf{x}_{i}$ & $i$th univariate signal of $\mathbf{X}$ & $\mathbf{C}$ & Connectivity matrix\\ 
	\hline
	$x_{i}(t)$ & $t$th sample of $\mathbf{x}_{i}$ & $c_{ij}$ & $ij$th entry of $\mathbf{C}$\\
	\hline
	$\Gamma$ & Graph-variate signal & $\theta$ & GVD connectivity \\	
	\hline
\end{tabular}
\end{table}
We seek to understand the general form of the connectivity proposed in \cite{Smit2016}, where the Dirichlet energy is taken as an instantaneous connectivity estimation of correlation in conjunction with an underlying graph constructed from Pearson's correlation coefficient. The motivation of which is threefold: i) in order to study if more suitable such forms exist, ii) in order to extend the method to different connectivity estimates, and iii) in order to study the general case where the graph is not constructed from the multivariate signal. Computationally, this generalisation is facilitated by formulating a tensor, $\underline{\mathbf{J}}\in\mathbb{R}^{n\times n\times p}$, whose elements are the output of a bivariate function, defined as
\begin{equation}\label{Jmat}
	J_{ijt} = \left\{
	\begin{array}{ccc}
	& F_{\mathcal{V}}(x_{i}(t),x_{j}(t)), & i\neq j\\
	& 0 , & i = j,\\
	\end{array}
	\right.
\end{equation}
for some function $F_{\mathcal{V}}$. Note that $F_{\mathcal{V}}$ is referred to as a node space function in the unified framework of multivariate signals and graphs, \eqref{eq4}, set forth in the Appendix. This is because it acts on the multivariate signal associated with the node set within $\Gamma$. We now proceed to define graph-variate signal analysis.
\begin{definition}
\textit{Graph-variate signal analysis} is the all-to-all bivariate analysis of the signal $\mathbf{X}$ weighted by the corresponding adjacency matrix $\mathbf{W}$ of the graph-variate signal $\Gamma$, taking the form
\begin{equation}\label{GV}
	(\mathbf{W}\circ \underline{\mathbf{J}}_{(t)})_{ij} = \left\{
	\begin{array}{ccc}
	& w_{ij}F_{\mathcal{V}}(x_{i}(t),x_{j}(t)), & i\neq j\\
	& 0 , & i = j,\\
	\end{array}
	\right.
\end{equation}
where $\underline{\mathbf{J}}_{(t)}$ denotes the $t$th $n\times n$ matrix of $\underline{\mathbf{J}}$ and $\circ$ is the Hadamard product. 
\end{definition}	

This way, a general node space function, $F_{\mathcal{V}}$, acting on $x_{i}$ and $x_{j}$ of the graph signal is weighted by $w_{ij}$, which encodes some measure of connectedness between nodes $i$ and $j$. This poses a new flexible analysis of multivariate signals embedded in a topology where the choice of $F_{\mathcal{V}}$ can be tailored to the given problem. 

It is important to note that this has not previously been considered in GSP. In the Appendix this is substantiated with the proposal of a general unified framework of multivariate signals and graphs. From this it is explained that methods in GSP tend to lie in the creation of functions of the graph adjacency matrix applied to the graph signal vector via normal matrix multiplication. Instead, graph-variate signal analysis is concerned with functions of the graph signal applied to the graph adjacency matrix using the Hadamard product. Proposition 1 then shows that the matrix multiplication of a graph adjacency matrix and graph signal vector can only encode linear bivariate functions of the graph signal samples without constants. Thus, graph-variate signal analysis is a conceptually and methodologically new framework for analysing multivariate signals using graphs.

\subsection{Graph-variate networks}
Interestingly, from \eqref{GV} we note that $\underline{\mathbf{\Delta}}_{(t)} = \mathbf{W}\circ \underline{\mathbf{J}}_{(t)}$ itself takes on a weighted adjacency matrix form and thus the tensor $\underline{\mathbf{\Delta}}\in\mathbb{R}^{n\times n\times p}$ is a multi-layer network of sequentially related graphs \cite{Dedom2013}. This is useful as we can then explore topological characteristics of a graph-variate signal at every sample. In classical network science, there are many methods proposed to analyse the topology of a graph by applying operations on the graph adjacency matrix \cite{Newm2010}. Such methods provide important insights and classifications of the interdependent relationships of the underlying objects. In our experiments, we will implement a simple example of a local clustering coefficient, $C_{loc}$ \cite{Watt1998}, of node $i$ at time $t$, defined for the graph-variate signal as	
\begin{equation}\label{clstr}
	C_{loc}(i,t) =  \sum_{j,k=1}^{n}\Delta_{ijt}\Delta_{ikt}\Delta_{jkt} = (\underline{\mathbf{\Delta}}_{(t)}^{3})_{ii}.
\end{equation}
This is computed for each node at each $t$  as the $i$\textsuperscript{th} diagonal element of the cube of $\underline{\mathbf{\Delta}}_{(t)}$. The reader is referred to e.g. \cite{Newm2010} for other possible network measures that could be used depending on the given problem.
	
\section{Graph-variate dynamic connectivity analysis}\label{tempcon}
Here, we shall focus on the special case in which the graph weights encode pairwise dependencies which have been estimated using the multivariate signal itself. The nomenclature of connectivity here is borrowed from neuroimaging \cite{Bull2009}, where large weights denote strong connectivity between two nodes and small weights denotes a lack of connectivity. The following makes use of the instantaneous amplitude and phase components of the analytic representation of the univariate signals $\mathbf{x}_{i}$, of $\mathbf{X}$, $x^{a}_{i}(t) = s^{a}_{i}(t)e^{j\phi_{i}}(t)$.

The connectivity between two nodes is generally established by a bivariate function of the signal pair. Doing this for all signal pairs of $\mathbf{X}$ establishes a weighted adjacency matrix of connectivities. We shall denote the connectivity adjacency matrix as $\mathbf{C}$, with entries $c_{ij}$ the connectivity between signals $i$ and $j$, to distinguish it from the general notation of an adjacency matrix, $\mathbf{W}$, where the edges need not necessarily be constructed from the signals. If the bivariate function is symmetric, then the matrix $\mathbf{C}$ is regarded as the weighted adjacency matrix of an undirected graph. Otherwise the graph is directed. We focus only on the undirected case in this article, however directed graphs may also be considered. To get a robust instantaneous measure of connectivity, we propose to filter an instantaneous function reflecting the formula of $\mathbf{C}$ by the stable dependencies of $\mathbf{C}$ itself. We do this in order to concentrate the instantaneous estimate on those connections known to exist and suppress those connections from which there is not enough evidence to suggest a true dependency.

Thus, we define GVD connectivity as a graph-variate signal analysis in which $\mathbf{W}$ = $\mathbf{C} \text{ of } \Gamma$ is an adjacency matrix of connectivities constructed from $\mathbf{X} \text{ of } \Gamma$ and $F_{\mathcal{V}}$ is a node space function acting as an instantaneous form of the bivariate function used to construct $\mathbf{C}$. This takes the form
\begin{equation}\label{tempconfun}
	\theta(\mathbf{x}_{i},\mathbf{x}_{j},t) = \left\{
	\begin{array}{cc}
	c_{ij}F_{\mathcal{V}}(x_{i}(t),x_{j}(t)), & i \neq j\\
	0, & i = j.
	\end{array}
	\right.
\end{equation}
Note that graph-variate signal analysis does allow the possibility to consider any possible bivariate function for any given graph, but caution is advised as this may lead to data dredging.
	
A particularly useful analysis for exploring the GVD connectivity associated with a particular node is the \textit{node GVD connectivity} 
\begin{equation}\label{nodetempcon}
	\theta_{i}(\mathbf{X},t) = \sum_{j = 1}^{n}c_{ij}F_{\mathcal{V}}(x_{i}(t),x_{j}(t)).
\end{equation}
We will use this a number of times in our experiments. The operator which extracts the vector of node temporal connectivities is defined in \eqref{diamond}.
	
Here we present node functions for three pertinent examples of connectivity adjacency matrices-- correlation, coherence and phase-lag index. For clarity of exposition, in each case we will first present the formulae for these connectivity estimates before going on to describe the chosen node space functions to compute GVD connectivity.
	
\subsection{Correlation}
Taking the connectivity estimate as the correlation coefficient, we have
\begin{equation}\label{corr}
	c_{ij} = \frac{\sum_{t\in T}(x_{i}(t)-\bar{\mathbf{x}}_{i})(x_{j}(t)-\bar{\mathbf{x}}_{j})}{\sqrt{\sum_{t\in T}(x_{i}(t)-\bar{\mathbf{x}}_{i})^{2}}\sqrt{\sum_{t}(x_{j}(t)-\bar{\mathbf{x}}_{j})^{2}}}
\end{equation}
where $T$ is the epoch of interest and $\bar{\mathbf{x}}_{i}$ is the mean of the values over time of the node $i$. In the preliminary formulation, Smith et al. \cite{Smit2016} presented dynamic connectivity as: 
\begin{equation}\label{sqd}
	\theta(\mathbf{x}_{i},\mathbf{x}_{j},t) = c_{ij}(\tilde{x}_{i}(t) - \tilde{x}_{j}(t))^{2},
\end{equation}
derived from the Dirichlet energy form from GSP \cite{Shum2013}. Here, $\tilde{\mathbf{x}}_{i}(t)$ is the normalised signal over the node space, i.e.
\begin{equation}
	\tilde{x}_{i}(t) = \frac{x_{i}(t) - \bar{\mathbf{x}}(t)}{\sqrt{\frac{1}{n-1}\sum_{k = 1}^{n}(x_{k}(t) - \bar{\mathbf{x}}(t))^{2}}},
\end{equation}
where $\bar{\mathbf{x}}(t) = \frac{1}{n}\sum_{k = 1}^{n}x_{k}(t)$ is the mean over nodes of the signal at time $t$. Notably, the entries of the matrix may be negative which is an important principle, as noted in \cite{Smit2016}, for maintaining the anti-correlative information. Differences in amplitudes at time $t$ reflect the amplitude dependent correlation coefficient \eqref{corr}. Small instantaneous differences reflect positive correlation and large instantaneous differences reflect negative correlation. 

However, correlation concerns the shape of signals with respect to their means \eqref{corr}. An instantaneous correlation may more usefully reflect this principle. Thus, we consider a function deriving more directly from \eqref{corr}:
\begin{equation}\label{icor}
	\theta(\mathbf{x}_{i},\mathbf{x}_{j},t) = \rho_{t}(i,j) = c_{ij}|(x_{i}(t)-\bar{\mathbf{x}}_{i})(x_{j}(t)-\bar{\mathbf{x}}_{j})|,
\end{equation}
where $\rho_{t}(i,j)$ is the shorthand which will be used at various points in the experiments. We shall compare \eqref{sqd} and \eqref{icor} in simulations and real data to help reveal the benefits of using this more relatable function than the squared difference coming arbitrarily from the Dirichlet form.
	
\subsection{Coherence}
The coherence of two nodes is a function of frequency, $\omega$, and can be interpreted as a correlation of signal components at $\omega$. For a chosen frequency band we thus have
\begin{equation}
	c_{ij} = \sum_{\omega\in\Omega}\frac{|P_{\mathbf{x}_{i}\mathbf{x}_{j}}(\omega)|^{2}}{P_{\mathbf{x}_{i}\mathbf{x}_{i}}(\omega)P_{\mathbf{x}_{j}\mathbf{x}_{j}}(\omega)},
\end{equation}
where $\Omega$ is a frequency band of interest, $P_{\mathbf{x}_{i}\mathbf{x}_{j}}$ is the cross-spectral density function of $\mathbf{x}_{i}$ and $\mathbf{x}_{j}$ and  $P_{\mathbf{x}_{i}\mathbf{x}_{i}}$ and  $P_{\mathbf{x}_{j}\mathbf{x}_{j}}$ the respective power spectral density functions \cite{Rapp1996}.

An instantaneous version of coherence should reflect the correlation of power of the two signals within a given frequency band. Thus, the first step is to bandpass the signal in the frequency of interest. After this, we take the envelope of the signals and look at their instantaneous correlations based on \eqref{sqd} and \eqref{icor}. That is, we consider
\begin{equation}\label{iamp}
	\theta(\mathbf{x}_{i},\mathbf{x}_{j},t) = c_{ij}(s^{a}_{i}(t)-s^{a}_{j}(t))^2.
\end{equation}
and
\begin{equation}\label{iampcor}
	\theta(\mathbf{x}_{i},\mathbf{x}_{j},t) = c_{ij}|(s^{a}_{i}(t)- \bar{\mathbf{s}}^{a}_{i})(s^{a}_{j}(t) - \bar{\mathbf{s}}^{a}_{j})|,
\end{equation}
respectively, as GVD connectivity estimates of instantaneous coherence.
	
\subsection{Phase-lag index}
The Phase-Lag Index (PLI) \cite{Stam2007} measures the consistent direction of phase differences between time-series, indicating lead/lag dependencies. As a connectivity estimate, we write
\begin{equation}\label{pli}
	c_{ij} = |\langle \text{sgn}(\phi_{i}(t)-\phi_{j}(t)) \rangle|,
\end{equation}
i.e. the magnitude of the average over time of the sign of differences of instantaneous phase, $\phi_i(t)$.
	
We choose $F_{\mathcal{V}}$ for phase-based connectivity indexes as the sign of the phase difference of the signals stemming directly from \eqref{pli}, giving
\begin{equation}\label{iphs}
	\theta(\mathbf{x}_{i},\mathbf{x}_{j},t) = c_{ij}\text{sgn}(\phi_{i}(t) - \phi_{j}(t)).
\end{equation}
Because of the negative symmetry of this function, the global GVD connectivity of the system at time $t$ is
\begin{equation}\label{phasesumzero}
	\begin{split}
	\sum_{i,j}\theta(\mathbf{x}_{i},\mathbf{x}_{j},t) &=\sum_{i<j}(\theta(\mathbf{x}_{i},\mathbf{x}_{j},t)+\theta(\mathbf{x}_{j},\mathbf{x}_{i},t))\\
	&=\sum_{i<j}(\theta(\mathbf{x}_{i},\mathbf{x}_{j},t)-\theta(\mathbf{x}_{i},\mathbf{x}_{j},t)) = 0.
	\end{split}
\end{equation}
However, we can sum over a subset of these elements to reveal the strength and general nature of the chosen elements to lead (positive) or lag (negative) in the network at the given epoch. Obvious choices for such a subset are singular nodes (summing over all $j$ for node $i$) or modules (summing over all $j$ for all $i$ in a given module).
	
\section{Experiments}
We will now apply the above outlined methods to several simulated and real datasets to provide document of their utility. An autoregressive model is implemented first to illustrate the broad idea and benefit of graph-variate signal analysis. We then extend this model to explore the ability of GVD connectivity to correctly discover differences between two large datasets which differ only by the presence (and lack thereof) of a single correlated couple. To test the effectiveness of temporal network clustering coefficient metric \eqref{clstr}, we devise a simple regime to detect a spheroid travelling over a 3D grid. We then apply our techniques to real high complexity datasets of geophysical well logs and EEG brain functional connectivity to provide evidence of the benefits delivered by a graph-variate analysis approach. Code of the simulated experiments and a function to produce graph-variate signal analysis is available at DOI 10.17605/OSF.IO/G82PV.
	
\subsection{Detecting correlated sources}
\begin{figure}[!tb]
	\centering
	\includegraphics[trim = 30 130 0 100,clip,scale=0.25]{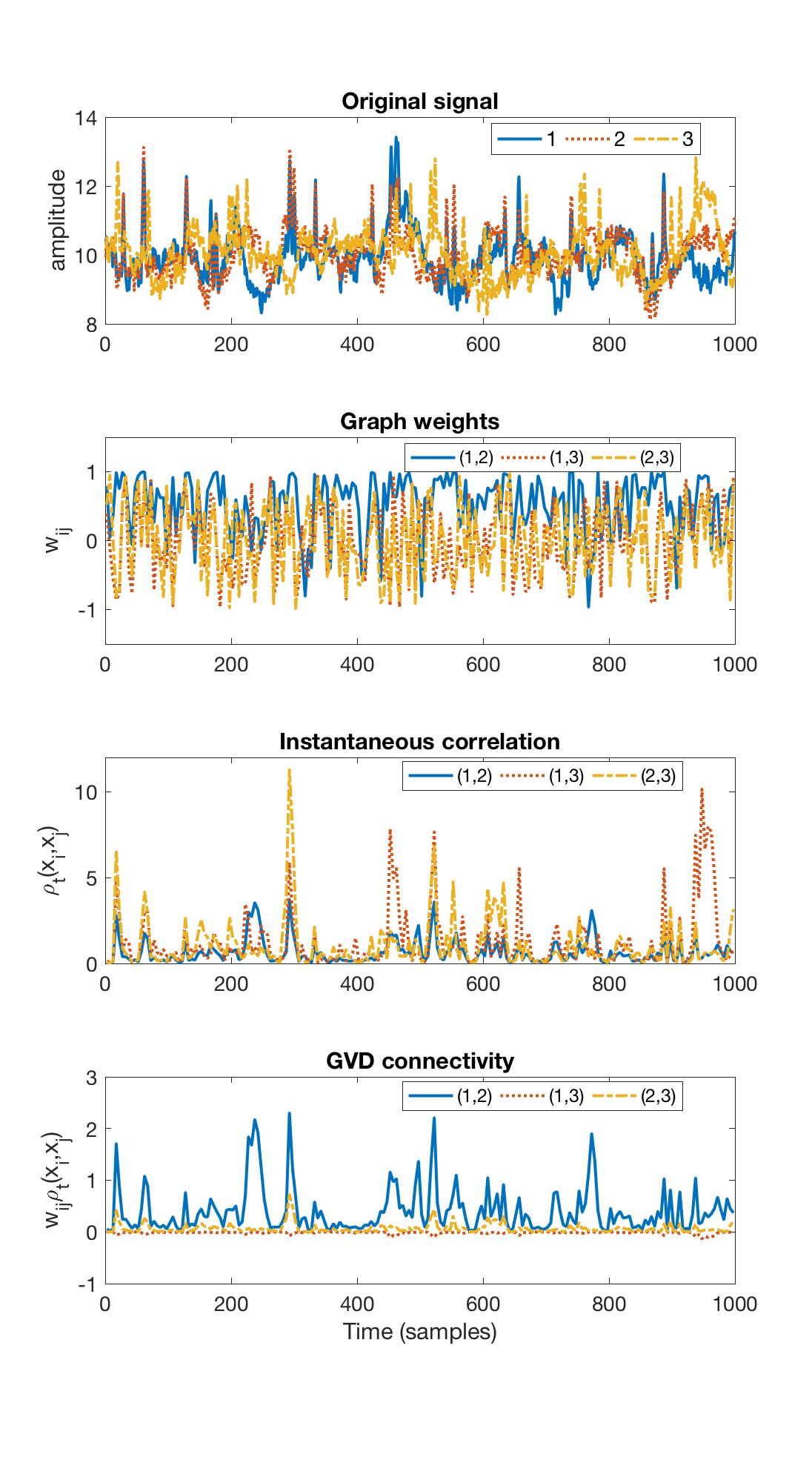}
	\caption{The original signal (top), dynamic graph weights (second), instantaneous correlations (third) and corresponding GVD connectivity (bottom) of edges as shown in the legend. The benefit of long-term graph weights is evident, where the GVD connectivity correctly emphasises important information (that related to edge (1,2)).}
	\label{toys}
\end{figure}

We generate 5 realisations, $1\times 1000$ vectors $\{\mathbf{z}_{i}\}_{i=1}^{5}$, of a stationary autoregressive process with governing equation
\begin{equation}\label{AR}
	z(t) = 0.5 + 0.7z(t-1) + 0.25z(t-2) + \epsilon,
\end{equation}
where $\epsilon\sim \mathcal{N}(0,0.1)$ and consider the multivariate signal
\begin{equation}
	\left[\begin{array}{c} \mathbf{x}_1 \\ \mathbf{x}_2 \\ \mathbf{x}_3 \end{array} \right] = \begin{matrix}\begin{bmatrix} \tfrac{1}{2} & \tfrac{1}{2} & 0 & 0 & 0 \\[2pt] \tfrac{1}{2} & 0 & \tfrac{1}{2} & 0 & 0 \\[2pt] 0 & 0 & 0 & \tfrac{1}{2} & \tfrac{1}{2} \end{bmatrix}\\\mbox{} \\ \mbox{} \end{matrix} \left[ \begin{array}{c} \mathbf{z}_1 \\[2pt] \mathbf{z}_2 \\[2pt] \mathbf{z}_3 \\[2pt] \mathbf{z}_4 \\[2pt] \mathbf{z}_5\end{array}\right],
\end{equation}
so that i) all $\mathbf{x}_{i}$ are the average of two realisations of \eqref{AR}, ii) $\mathbf{x}_{1}$ and $\mathbf{x}_{2}$ are correlated via the information in $\mathbf{z}_{1}$, and iii) $\mathbf{x}_{3}$ is independent of $\mathbf{x}_{1}$ and $\mathbf{x}_{2}$. Fig. \ref{toys} shows the computation of instantaneous correlation coefficients and corresponding node GVD connectivity computed using correlation coefficient \eqref{icor} over the entire signal. The corresponding graph weights are $c_{12} = 0.6934, c_{13} = -0.0576, c_{23} = 0.0943$. Node GVD connectivity (bottom) is computed over 5 samples in non-overlapping windows. The corresponding short-term graph weights (computed over 5 samples) and the un-weighted instantaneous correlation are shown in the second and third panels of Fig. \ref{toys}, respectively. Unsubstantiated dependencies are produced using the short-term graph weight and instantaneous correlation methods where often the three outputs are roughly equivalent. GVD connectivity, on the other hand suppresses the uncorrelated data using the long-term connectivity estimates and the prevailing information comes forth from the truly correlated data relating to edge $(1,2)$. This is most obviously seen in comparing instantaneous correlation (third panel) with GVD connectivity (bottom), where the signals are identical except that GVD connectivity weights them by long-term correlations. Hence the dash-dotted, $(2,3)$, and dotted, $(1,3)$, time-series are suppressed relative to the solid time-series, $(1,2)$.
	
\begin{figure}[!tb]
	\centering
	\includegraphics[trim = 150 0 0 0,clip,scale=0.125]{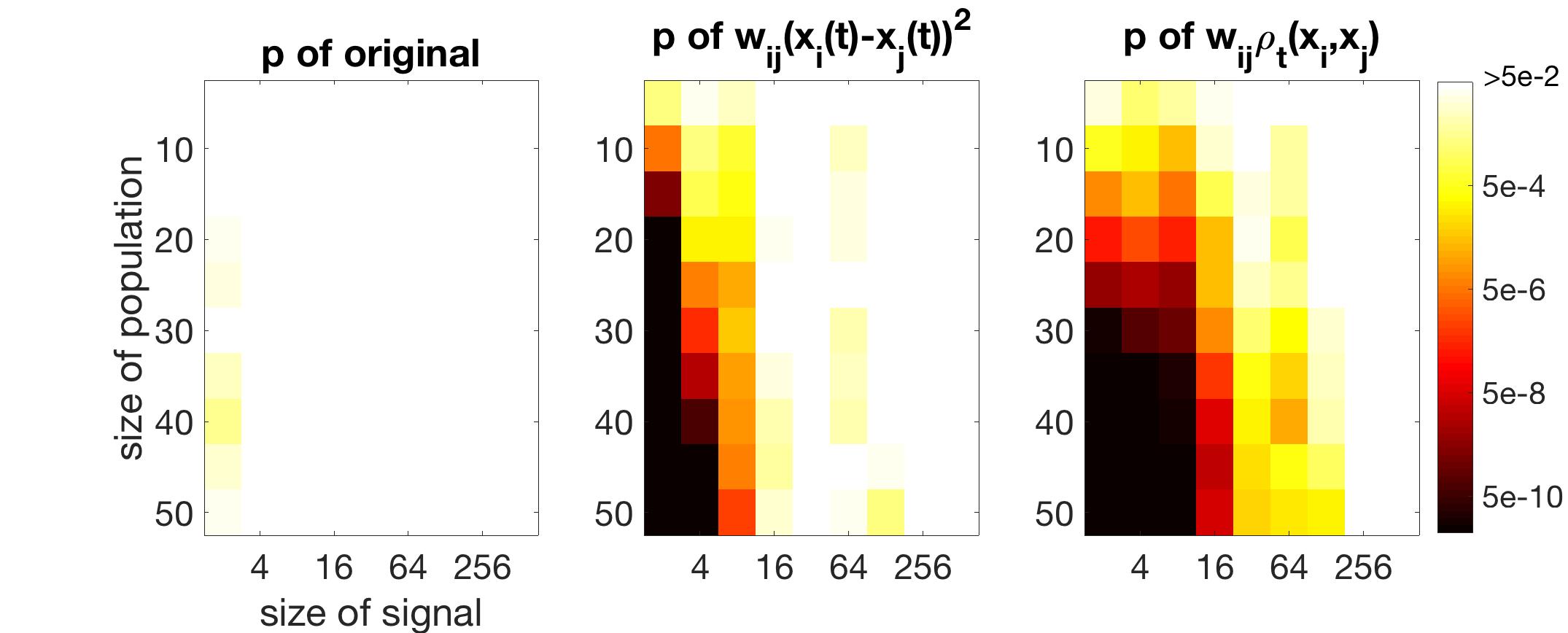}
	\caption{The $p$-values of one-sample $t$-tests for correctly identifying correlated sources for different sizes of multivariate signal ($x$-axis) and population ($y$-axis).  Shown for the original signal (left) and GVD node connectivity with squared difference (middle) and instantaneous correlation (right). White indicates a non-significant difference, black indicates a $p$-value value smaller than $5\times 10^{-10}$.}
	\label{corrsource}
\end{figure}

We now extend this to quantitatively assess the size of multivariate signal from which the presence of a single couple of correlated signals can be detected. Following the same autoregressive process as \eqref{AR}, we generate $2\times h$ realisations for $h = 2,4,8,16,\dots,512$. Two sets of signals are then formed. The first uncorrelated set takes the average of each consecutive disjoint couple of realisations as the multivariate signal $\mathbf{X}\in\mathbb{R}^{h\times 1000}$. The second set is almost the same except the first signal is formed from the first and third (rather than first and second) realisations so as to be correlated with the second signal. These two sets of signals can thus be formulated as
\begin{equation}
	 \begin{matrix}\begin{bmatrix} \mathbf{x}_1 \\ \mathbf{x}_2 \\ \mathbf{x}_3  \\ \vdots \\ \mathbf{x}_{h} \end{bmatrix}\\\mbox{} \\ \mbox{} \end{matrix} = \begin{matrix}\begin{bmatrix} \tfrac{1}{2} & \tfrac{1}{2} & 0 & 0 & 0 & 0 & \dots & 0 & 0 \\ 0 & 0 & \tfrac{1}{2} & \tfrac{1}{2} & 0 & 0 & \dots & 0 & 0 \\ 0 & 0 & 0 & 0 & \tfrac{1}{2} & \tfrac{1}{2} & \dots & 0 & 0 \\  \vdots & \vdots & \vdots & \vdots & &  & \ddots & \vdots & \vdots \\ 0 & 0 & 0 & 0 & 0 & 0 & \dots & \tfrac{1}{2} & \tfrac{1}{2} \end{bmatrix}\\\mbox{} \\ \mbox{} \end{matrix} \begin{matrix}\begin{bmatrix} \mathbf{z}_1 \\ \mathbf{z}_2 \\ \mathbf{z}_3 \\ \vdots \\ \mathbf{z}_{2h} \end{bmatrix}\\\mbox{} \\ \mbox{} \end{matrix},
\end{equation}
and
\begin{equation}
	 \begin{matrix}\begin{bmatrix} \mathbf{x}_1 \\ \mathbf{x}_2 \\ \mathbf{x}_3  \\ \vdots \\ \mathbf{x}_{h} \end{bmatrix}\\\mbox{} \\ \mbox{} \end{matrix} = \begin{matrix}\begin{bmatrix} \tfrac{1}{2} & 0 & \tfrac{1}{2} & 0 & 0 & 0 & \dots & 0 & 0 \\[2pt] 0 & 0 & \tfrac{1}{2} & \tfrac{1}{2} & 0 & 0 & \dots & 0 & 0 \\[2pt] 0 & 0 & 0 & 0 & \tfrac{1}{2} & \tfrac{1}{2} & \dots & 0 & 0 \\[2pt]  \vdots & \vdots & \vdots & \vdots & &  & \ddots & \vdots & \vdots \\[2pt] 0 & 0 & 0 & 0 & 0 & 0 & \dots & \tfrac{1}{2} & \tfrac{1}{2} \end{bmatrix}\\\mbox{} \\ \mbox{} \end{matrix} \begin{matrix}\begin{bmatrix} \mathbf{z}_1 \\[2pt] \mathbf{z}_2 \\[2pt] \mathbf{z}_3 \\[2pt] \vdots \\[2pt] \mathbf{z}_{2h}  \end{bmatrix}\\\mbox{} \\ \mbox{} \end{matrix}.
\end{equation}
We generate populations of such multivariate signals of sizes $5,10,15,\dots, 50$ to track effects due to population size. We compute the difference between the uncorrelated and correlated original signals alongside differences in GVD connectivity analyses using \eqref{sqd} and \eqref{icor} and sum over time. These are then put to a one-sample $t$-test on the null hypothesis that the population values have a zero mean, with significance indicating rejection of the null hypothesis at the $\alpha = 0.05$ level. The results for each population and signal size are shown in Fig. \ref{corrsource}.
	
The values for the original signals are provided for reference only. We do not expect them to perform well given that they rely only on signal magnitudes. The results clearly indicate that GVD connectivity using instantaneous correlation has greater sensitivity to differences than using squared difference. Specifically, we can state that GVD connectivity with instantaneous correlation can correctly and reliably detect the single correlated source from multivariate signals of size 128 across signal population sizes of 25 or more. While it is true that squared difference can detect differences in 128 signals with a population size of 50, this cannot be seen as reliable since it fails to detect any difference in the cases with signals of size 32.
		
\subsection{Spheroid travelling randomly on a 3D grid}
This problem studies a case where the graph is not constructed from the signal, but instead provides the geometry in which the signal is set. We construct a $10\times 10\times 10$ grid in Euclidean space where each point is associated with a univariate signal. Each pair of horizontally and vertically neighbouring grid points, $(i,j)$, are at distance $d_{ij}=1$ from each other. A weighted connectivity graph is then formed from the inverse distance, computed as $w_{ij} = \exp(-d_{ij}^{2}/4)$. The signal to study is created following the pseudocode:

\begin{algorithm}
\caption{Generate spheroid signal}\label{sphalg}
\begin{algorithmic}[1]
\State Initiate signal $\mathbf{X}\in\mathbb{R}^{n\times 1000}$ with entries $X_{ij} \sim\mathcal{N}(0,0.3)$
\State At $t=1$ choose spheroid centre, $s(1)$, randomly from integers up to $n$
\State Set $X_{s(1)1} = X_{s(1)1} + \delta$ 
\State Set $X_{z1} = X_{z1} + \tfrac{3}{4}\delta$ where $z$ s.t. $w_{s(1)z} = \exp(-1^{2}/4)$
\For{$t=2$ up to $1000$}
\State Choose $s(t)$ randomly from $z$  s.t. $w_{s(t-1)z} = \exp(-1^{2}/4)$
\State Set $X_{st} = X_{st} + \delta$ 
\State Set $X_{zt} = X_{zt} + \tfrac{3}{4}\delta$ where $z$ s.t. $w_{sz} = \exp(-1^{2}/4)$
\EndFor
\end{algorithmic}
\end{algorithm}

We can liken this to a spheroid travelling at random over a grainy image where at each time point the spheroid moves randomly to a neighbouring node on the grid. This process is implemented for values of $\delta$ ranging from 0.1 in steps of 0.1 up to 0.9. We now consider the appropriate node space function to use in this scenario. The randomness of movement means that using approaches which try to assess a direction, such as Kalman filtering, are of little value. Thus, a more basic maximisation approach is adopted. We implement graph-variate signal analyses using a multi-layer graph, $\underline{\mathbf{\Delta}}$, where each layer relates to the graph-variate signal at time sample $t$. Considering that higher amplitudes close together should produce high values, we choose a node space function for $\mathbf{\underline{\Delta}}$ which takes the average of each signal pair, so that:
\begin{equation}
	\Delta_{ijt} = \tfrac{1}{2}w_{ij}(x_{i}(t)+x_{j}(t)).
\end{equation}
We then calculate the weighted clustering coefficient, $C_{loc}$, from \eqref{clstr}, respectively, at each node at each point in time. The task is then to detect the spheroid. Alongside a simple comparison against the node with highest amplitude, a number of graph filtering approaches are implemented. We compare with the graph adjacency matrix with self-loops, $\widehat{\mathbf{W}} = \mathbf{I} + \mathbf{W}$ \cite{Sand2013}, and the graph Laplacian \cite{Shum2013}, aswell as using the heat kernel, $e^{-\mathbf{\tau L}}$ \cite{Shum2013}. That is, at time $t$, we select the highest value of the vectors $\widehat{\mathbf{W}}\mathbf{X}(t)$, $\mathbf{L}\mathbf{X}(t)$, and $e^{-\mathbf{L}}\mathbf{X}(t)$. To align with the notion of the clustering coefficient approach \eqref{clstr}, we also look at the cubed versions $\widehat{\mathbf{W}}^{3}\mathbf{X}(t)$, $\mathbf{L}^{3}\mathbf{X}(t)$, and $e^{-3\mathbf{L}}\mathbf{X}(t)$. The cube of the graph adjacency matrix contains all paths of length 3 between nodes $i$ and $j$ at each $ij$th entry (i.e. the number of triangles of which $(i,j)$ is an edge).
		
\begin{figure}[!tb]
	\centering
	\includegraphics[trim = 10 20 0 0, scale = 0.3]{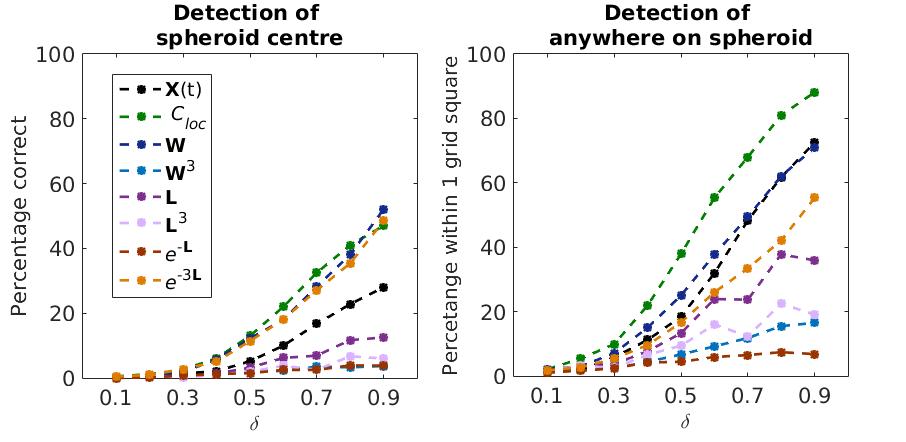}
	\caption{Total of correct estimations, left, and estimations corresponding to anywhere on the spheroid, right, out of 1000 time points using amplitude height only, $\mathbf{X}(t)$, graph-variate clustering coefficient, $C_{loc}$, and GSP filtering approaches. Plotted against $\delta$-- the increased amplitude given to the central point of the spheroid.}
	\label{clusterRe}
\end{figure}

Fig. \ref{clusterRe} details the number of correctly identified spheroid centres from the largest values obtained by each approach (left) and the number of identifications at any point of the spheroid (right), i.e. within one grid square of centre. Our approach using $C_{loc}$ (green) achieves best results in 7/9 values of $\delta$ in the former and in all cases in the latter. It also shows best overall results, see Table \ref{ClstrTable}. It is one percentage point clear of the next best in detecting the centre and nearly ten percentage points clear of the next best in detecting any part of the spheroid. Of the GSP approaches, the best are the single adjacency matrix filter $\widehat{\mathbf{W}}$ (Fig. \ref{clusterRe}, dark blue) and the heat kernel $e^{-3\mathbf{L}}$ (orange), which perform relatively well in detecting the centre point. However, they fair much less well when taking into account the sides of the spheroid. Indeed, in this instance, they do not fair much better than the default maximum amplitude approach (black). Since $\mathbf{W}$ and $\mathbf{L}$ fair better than their cubic versions, we know that the improvement noted by the clustering coefficient method is not down to the cube of the graph distance information resulting from \eqref{clstr}. Instead, it is the combination of the graph and signal information which leads to increased accuracy.

\begin{table}[!tb]
	\caption{Percentages (\%) for different methods in correctly locating spheroid centre (Centre) and in identifying spheroid at any point (Any) over all sizes of $\delta$}
	\label{ClstrTable}
	\centering
	\begin{tabular}{|c|c|c|c|c|c|c|c|c|}
		\hline
		Locate & max & $C_{loc}$ & $\hat{\mathbf{W}}$ & $\hat{\mathbf{W}}^{3}$ & $\mathbf{L}$ & $\mathbf{L}^{3}$ & $e^{-\mathbf{L}}$ & $e^{-3\mathbf{L}}$ \\
		\hline
		Centre & 9.7 & \underline{18.4} & 17.4 & 1.9 & 4.8 & 2.5 & 1.8 & 16.6\\
		Any & 28.2 & \underline{41.1} & 30.3 & 7.7 & 16.7 & 10.3 & 4.5 & 21.4\\
		\hline
	\end{tabular}	
\end{table}
		
\begin{figure}[!tb]
	\centering
	\includegraphics[trim = 0 40 0 10,clip,scale = 0.23]{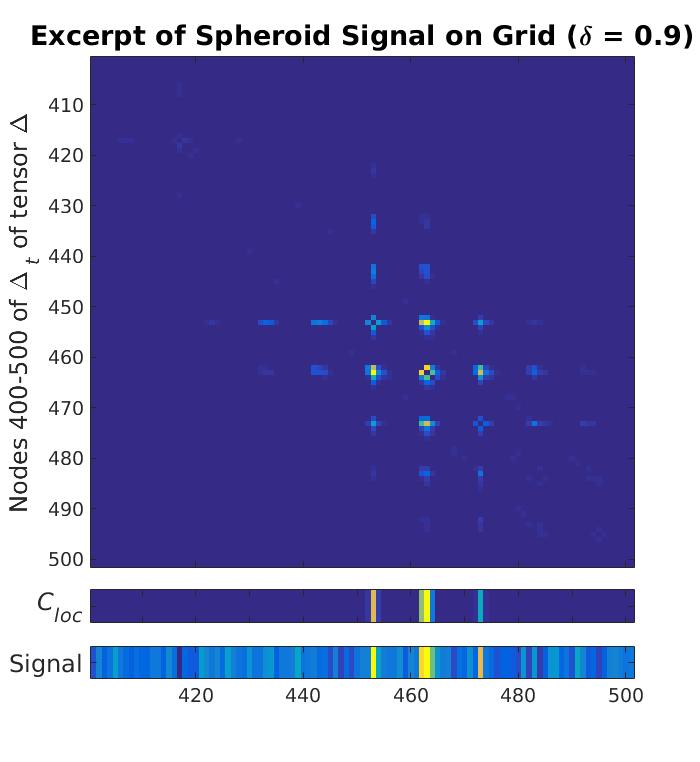}
	\caption{Example of reduced noise and increased accuracy through clustering coefficient in the spheroid detection problem. The highest amplitude is detected at node 452, however the maximum clustering coefficient, $C_{loc}$, detects the actual centre at node 462.}
	\label{clusterEx}
\end{figure}

An example of how the proposed method is able to correctly identify a spheroid centre which is not picked up using the highest amplitude alone is shown in Fig. \ref{clusterEx}. In this example, the increased amplitude of $3/4\delta$ given to one of the nearest nodes, 452, provides a larger overall amplitude to the $\delta$ given to the central node. By using the graph-variate method, however, this error due to noise is corrected since most of the nearest nodes to 452 have a very small comparative amplitude to those of the true centre at 462.
		
Analysis of the formulation of the $C_{loc}$ shows its power for the suppression of noise and promotion of clustered phenomena. In the problem illustrated we can consider the expected value of the signal triple 
\begin{align}
&E[(x_{i}+x_{j})(x_{j}+x_{k})(x_{k}+x_{i})] = 8E[x^{3}] \notag\\
&= 8(\mu_{3} + 3E[X]E[X^{2}] + 2(E[X])^{3}) \notag\\
&= 8(\mu_{3} + 3\mu(\sigma^{2}+\mu^{2}) + 2\mu^{3}),
\end{align}
where $\mu$ is the mean and $\mu_{3}$ is the third moment of variable $x$. For only noisy data $x\sim\mathcal{N}(0,\sigma^{2})$, this is just zero from the fact that odd moments of a symmetric distribution are zero and $\mu = 0$. On the other hand, the expected value for $x\sim\mathcal{N}(\delta,\sigma^{2})$, i.e. data with true value $\delta$ in the presence of noise, is  $24\sigma^2\delta + 40\delta^{3}$. In the GSP filtering approaches, the adjacency matrix provides $E[x_{i}] = 0$ for noise and $E[x_{i}] = \delta$ for the true value. This explains why it also fairs well at detecting the correct centre point. The Laplacian, on the other hand, provides $E[x_{i}-x_{j}]$ which is zero for both noise and true value, explaining its poor performance here. We note that this experiment may be too specific to provide a general sense of these approaches. However, this highlights the necessity for the appropriate consideration of analysis for the problem at hand, which can be assessed more fully within the proposed unified framework in the Appendix. To increase comparability, and in the pursuit of a simple example, these approaches are chosen to be free from parameters and more complicated methodologies such as using iterative denoising. We recognise, though, that more elaborate complementary methodologies such as implementing wavelets using a dictionary of spheroid shaped atoms \cite{Shuman2015} or joint time-graph denoising \cite{Grassi2017} may provide a more intensive treatment of the problem.
 	
\subsection{Gammay ray radiation from well logs}
Signals of gamma ray radiation measured in API (American Petroleum Institute) corrected gamma counts across several kilometres (one sample per metre) underground were acquired from well logs in the Kansas Geological survey \cite{KansGeo2017}. Note, these signals are sampled with respect to distance underground (depth) rather than time. Gamma ray radiation is recorded in order to detect shale (indicated by greater Gamma radiation) and is thus useful in oil discovery \cite{Rider2011}. We collected data for the month of June. As of the 24th of June 2017, data from 23 sites had been uploaded. Of these, one site had no gamma ray data and one pair of duplicate data were found. Each remaining site had one univariate signal of gamma ray counts in API sampled at each meter underground, with respect to sea level.  These were recorded over various depths. Visual inspection was used to assess a suitable epoch with recordings for as many signals as possible. This was found between 2-4km with 17 signals. Large correlation coefficients of these signals would indicate similarity in the geology. Thus, graph-variate signal analysis should be able to detect substantial changes in geology over large distances in a quick and easy way. We compute the correlation coefficient adjacency matrices \eqref{corr} and compare squared difference \eqref{sqd} and instantaneous correlation \eqref{icor} node functions in a GVD connectivity analysis of the gamma ray signals.
\begin{figure}[!t]
	\centering
	\includegraphics[trim = 25 20 0 20,clip,scale= 0.3]{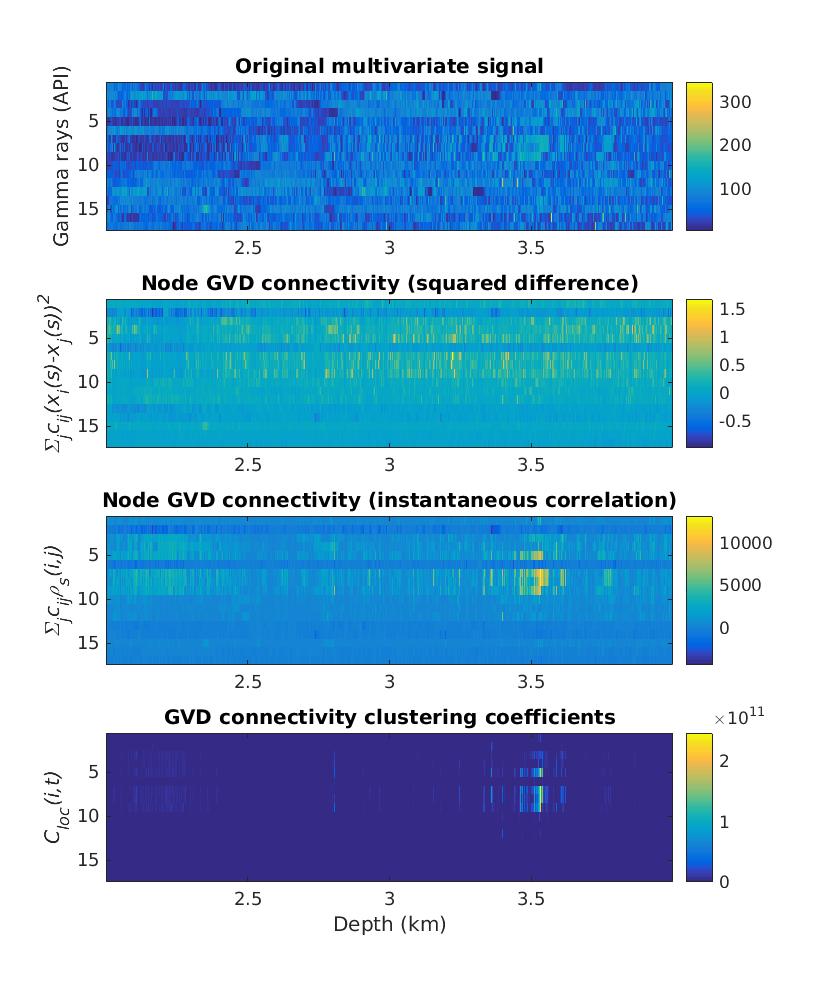}
	\caption{Analysis of seismic data from the June Kansas Geological Survey. Top shows the original gamma ray velocity data (American Petroleum Institute (API) units) taken at different 17 different locations, plotted against depth underground. Second and third plots show the node GVD connectivity using correlation graphs with squared difference and instantaneous correlation node functions, respectively. Bottom shows the clustering coefficients at each node of GVD connectivity of instantaneous correlation.}
	\label{seismic}
\end{figure}
	
Fig. \ref{seismic} shows the results of these versions of GVD connectivity (second and third panels, respectively) alongside the original signal (top). Generally, we look for gamma ray counts which are sustained over many tens of metres, indicative of possible reservoirs \cite{Rider2011}. Immediate observation of the data shows how GVD connectivity aids manual scrutiny of the signal (left) by dramatically reducing activity, particularly for instantaneous correlation. This makes it easy to spot some immediate epochs and signals of interest. GVD connectivity with instantaneous correlation shows up some very interesting sustained activity occurring between 3.375-3.4km across several signals. Positive and large activity here suggests that these signals are recorded at similar locations and thus that the related activity indicates a shale component at this depth covering the ground between these sites. Using the graph-variate clustering coefficient \eqref{clstr}, this activity clearly sticks out as the most significant correlated activity here (Fig. \ref{seismic}, bottom), where the rest of the activity being sent close to zero indicates uncorrelated and/or noisy data.
	
\subsection{GVD connectivity of EEG data}\label{EEGdata}
In this experiment we study an eyes-closed, eyes-open dataset of 129-channel EEG activity. This dataset is available online under an open database license from the Neurophysiological Biomarker Toolbox tutorial \cite{NBTdata}. It consists of data for 16 volunteers and is down-sampled to 200Hz. We used the clean dataset which we re-referenced to an average reference  and filtered in the Alpha band (8-13Hz) before further analysis. Alpha activity is well known to undergo notable changes between these states \cite{Barr2007}, thus such a dataset provides a solid testing ground for the use of our techniques on complex brain recordings.
	
\begin{figure}[!t]
	\centering
	\includegraphics[trim = 0 20 40 10,clip,scale=0.37]{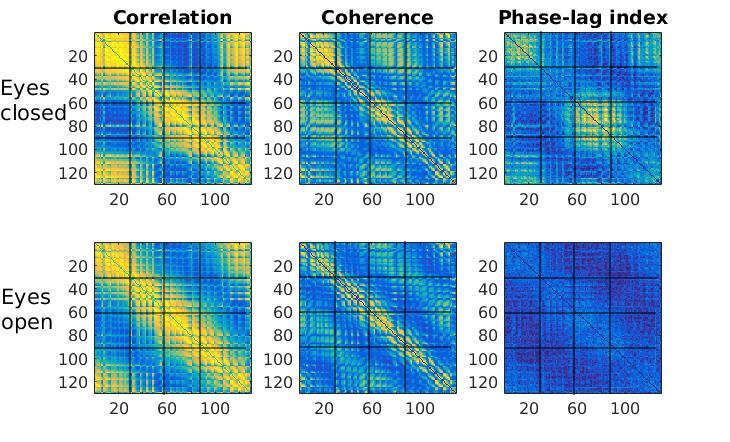}
	\caption{Weighted graph adjacency matrices of correlation, coherence and PLI for eyes closed (top) and eyes open (bottom) conditions. The colour axes, light/yellow being the largest weights, are the same for eyes open and eyes closed conditions. Modules are selected based on the most different activity between conditions- Module 1: nodes 1-30, Module 2: nodes 60-90, indicated by the black lines.}
	\label{meangraphs}
\end{figure}

The recordings are long-- 4.4355 $\pm$ 0.2861 mins (mean $\pm$ standard deviation)-- allowing us to take windows starting at the 1000\textsuperscript{th} sample (5s), to avoid the possibility of pre-processing artefacts at the beginning of the signal. We choose epochs, $\tau$, lasting $16,32,64,\dots, 2048$ samples (80ms up to 10.24s). We then investigate dynamic connectivity using correlation, coherence and PLI in Alpha within each epoch $\tau$. For analysis, modules (subsets of nodes) of interest are chosen based on observable differences in the average weights over graphs computed from the largest window ($\tau = 2048$), Fig. \ref{meangraphs}. Choosing modules, instead of global connectivity, allows us to compute our phase-based methods without redundancy \eqref{phasesumzero}. Clearly, around 1-30 nodes and 60-90 nodes show differences in all connectivity measures (Fig. \ref{meangraphs}, black lines mark nodes 30, 60 and 90), thus we choose these as Module A and Module B, respectively. Modular connectivity is computed, following the formula for modular Dirichlet energy in \cite{Smit2016}, as:
\begin{equation}\label{modular}
	\frac{1}{|T|}\sum_{t\in T}\sum_{i\in\mathcal{V}_{x}}\sum_{j\in \mathcal{V}}c_{ij}F_{\mathcal{V}}(x_{i}(t),x_{j}(t)),
\end{equation}
where $\mathcal{V}_{x}$ are the module nodes and $T$ is an epoch of interest within $\tau$. Here, $i$ sums over the module and $j$ sums over the entire graph to assess the modules effects on the entire graph. Equation \eqref{modular} is applied for correlation using \eqref{sqd} and \eqref{icor}, coherence using \eqref{iamp} and \eqref{iampcor}, and PLI using \eqref{iphs}.
	
For this dataset we seek to clarify the usefulness of our methods compared to weighted graphs by themselves, as implemented in e.g. \cite{Doron2012,Braun2015,Braun2016,Leonardi2015}, as well as the benefit of the graph support in GVD connectivity as opposed to using un-weighted node space functions i.e. putting all $c_{ij} = 1$ in \eqref{modular}. In the latter case, efforts similar to this have been made at determining dynamic connectivity using instantaneous phase differences in fMRI \cite{Glerean2012}.
	
\begin{figure}[!t]
	\centering
	\includegraphics[trim = 0 20 0 20,clip,scale=0.325]{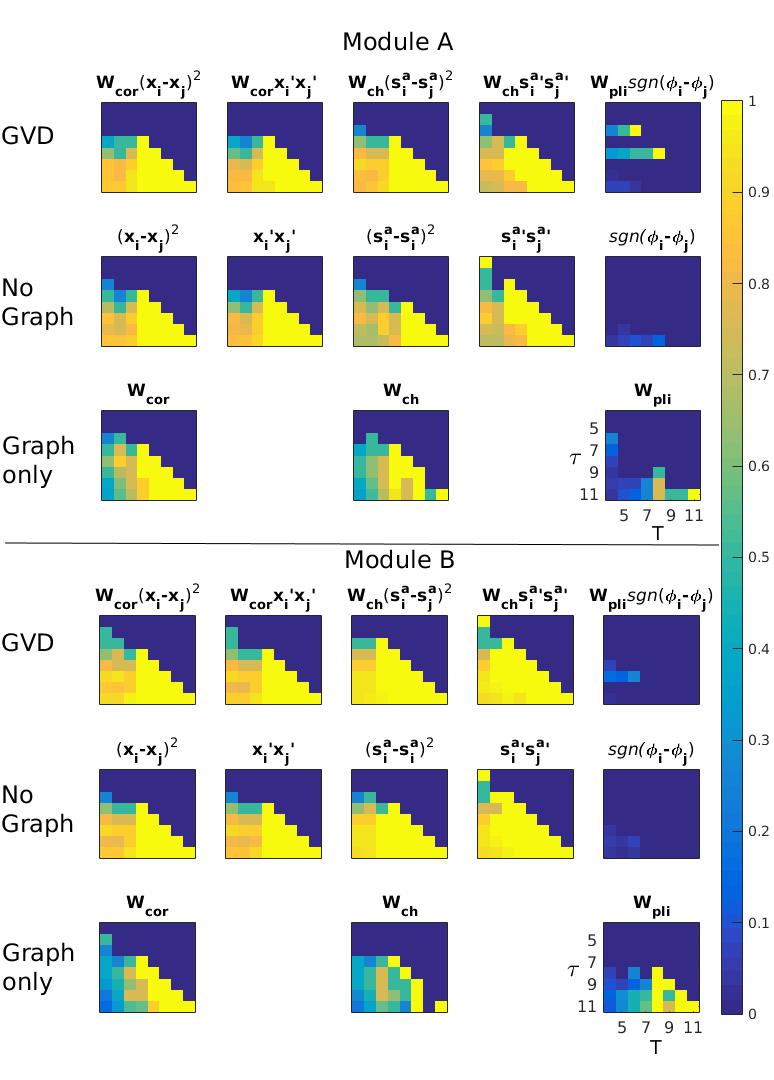}
	\caption{ Results of eyes open vs eyes closed EEG data for Module A, top and Module B, bottom, plotted by density of $p$ values which are significant for $T<\tau$. GVD (first row) is GVD connectivity where the graph comes from $\tau$ and the GVD is computed over $T$. The axes of $\tau$ against $T$, shown on the bottom right plot, indicates the signal length considered in powers of 2, i.e. 5 is $2^5 = 32$, etc. No graph (second row) is the non-graph weighted node space function. Graph only (third row) refers to graphs computed over $T$. $\mathbf{W}_{cor}$ is the adjacency matrix of correlations, $\mathbf{W}_{ch}$ of coherence and $\mathbf{W}_{pli}$ of PLI. Here, $\mathbf{x}_{i}$ is the original signal, $s_{i}^{a}$ the signal envelope and $\phi_{i}$ the instantaneous phase, where $\mathbf{x}_{i}'$ and $\mathbf{s}^{a}_{i}$$'$ are the signals minus their expected values as in \eqref{icor} and \eqref{iampcor}, respectively.}
	\label{FrontMu}
\end{figure}

For GVD connectivity let $\tau$ refer	to the epoch used to compute the long-term graph weight, $c_{ij}$, and let $T$ refer to the length of disjoint windows within $\tau$ used to compute the average of the instantaneous node function, $F_{\mathcal{V}}$ in \eqref{modular}. For modules A and B, we compute GVD connectivity over the epoch pair $(T,\tau)$ such that $T\leq \tau \in \{16,32,64,\dots,2048\}$. This gives a total of 36 cases corresponding to each combination of $(T,\tau)$ with a minimum of one $p$-value (when $T=\tau$) and maximum of $\tau/T$ = 2048/16 = 128 $p$-values in each case. For each $(T,\tau)$ we then compute the density (number of differences found out of total possible differences) of significant $p$-values from paired $t$-tests of eyes closed vs eyes open conditions across the 16 participants. The results for each $(T,\tau)$ are shown in Fig. \ref{FrontMu} for modules A and B for GVD connectivity, the node functions by themselves (No graph) and a dynamic graph approach (Graph only).

It is clear for both modules that the GVD connectivity approach performs better than the standard dynamic connectivity approach for correlation and coherence. The phase-lag index fairs poorly in this paradigm in general, but we shall see later that it may be leveraged to greater effect in time-locked task presentation data. It is not clear from observation if the GVD connectivity approach is better than the node space functions alone (No graph). To see this more evidently, we compute i) the number of cases, $(T,\tau)$, for which GVD connectivity outperforms the no graph approach and vice versa, and ii) the greater number of significant $p$-values shown by GVD connectivity within those cases and vice versa. Table \ref{EEGTable1} shows the results. 

We see that GVD connectivity consistently outperforms the node function by itself. There are a total of 45 cases (consisting of 118 $p$-values) in which it exceeds the node function alone in module A, and 28 cases (consisting of 50 $p$-values) in which it exceeds the node function alone in module B. The opposite, in which the node function alone exceeds GVD connectivity is much lower with just 16 cases (consisting of 17 $p$-values) in module A, and 10 cases (consisting of 13 $p$-values) in module B.  

\begin{table}[!t]
	\caption{Number of cases $(T,\tau)$ and $p$-values within those cases (cases:$p$-values) for which GVD connectivity (GVD) finds more significant differences ($>$) than node functions alone (NF) and vice versa. First column indicates GVD method used (graph/node function) where Cor- correlation, Ch- Coherence, sqd- squared difference, ico- instantaneous correlation and phs- sign of phase difference.}
	\label{EEGTable1}
	\centering
	\begin{tabular}{|c|c|c|c|c|} 
	\hline
	\textbf{Method} & \multicolumn{2}{c|}{\textbf{Module A}} & \multicolumn{2}{c|}{\textbf{Module B}}\\
	\hline
	| & GVD$>$NF & NF$>$GVD & GVD$>$NF & NF$>$GVD\\
	\hline
	Cor/sqd & 8:16  & 3:3 & 10:21 & 0:0\\
	Cor/ico & 4:4   & 3:3 &	6:11  & 0:0\\ 
	Ch/sqd  & 14:62	& 0:0 &	6:7   & 1:1\\
	Ch/ico  & 9:15	& 6:6 &	2:2   & 3:3\\
	PLI/phs	& 10:21	& 4:5 &	4:9	  & 6:9\\
	\hline
	Total	& 45:118 & 16:17 & 28:50 & 10:13\\
		\hline
	\end{tabular}	
\end{table}

To try the PLI in a more appropriate task-related setting where consistent phase dependencies of brain function over many trials can be picked out, we look at a face presentation task detailed in \cite{Henson2011}. The dataset consists of 16 subjects undergoing a face presentation task lasting 1.5 seconds (0.5s pre-stimulus) downsampled from 1kHz to 250Hz. Mean and standard deviation of trials is 294.19 $\pm$ 2.32. After bandpassing in Alpha (8-13Hz), the PLI is computed for each trial and then averaged over trials to construct an adjacency matrix per subject. Graph-variate analysis with and without the weighted adjacency matrix is then conducted using the sign of instantaneous phase differences. This is conducted per trial and then averaged over trials, after which the absolute value is taken. 

\begin{figure}[!t]
	\centering
	\includegraphics[trim = 10 50 0 20,clip,scale=0.45]{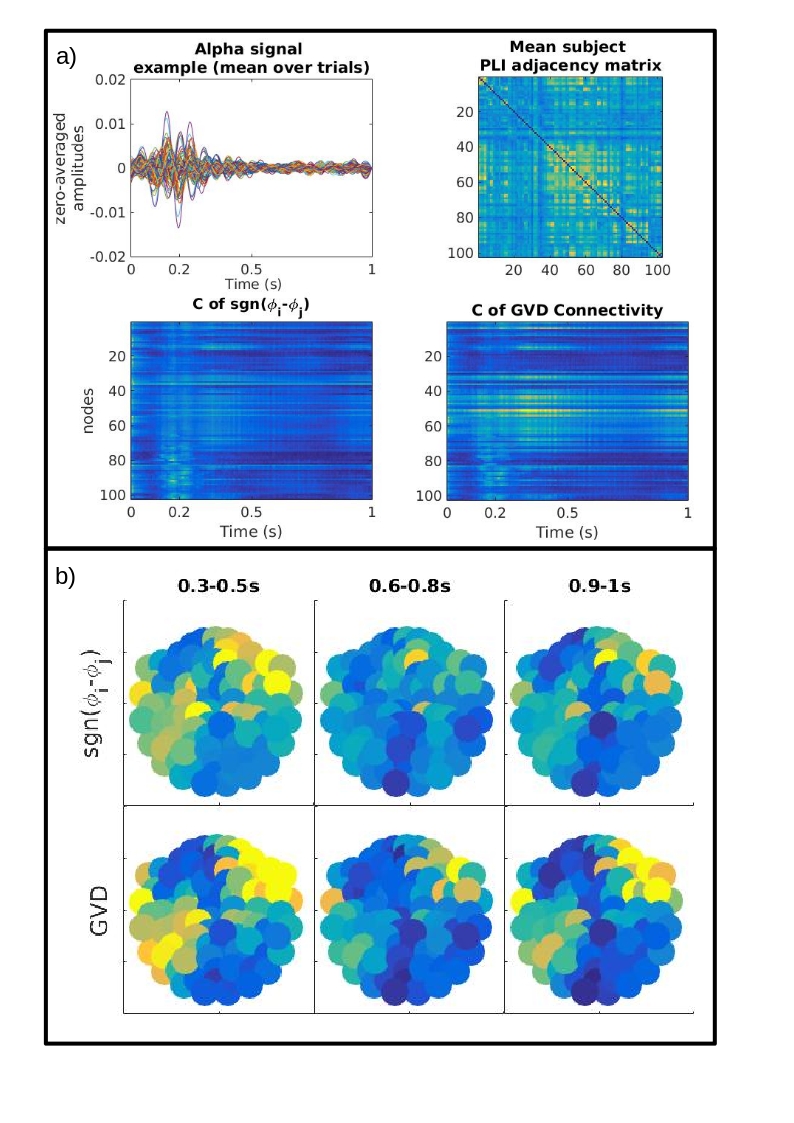}
	\caption{a) Phase activity from a face presentation task. Top left is the alpha signal for one subject. Top right is the mean connectivity adjacency matrix over all subjects. Bottom left is $C_{loc}$ for each node at each time point for instantaneous PLI, averaged over subjects. Bottom right is $C_{loc}$ for each node at each time point for GVD connectivity, averaged over subjects. b) Topoplots of the sum of $C_{loc}$ phase activity in a given time window from a face presentation task. Colour axis has a minimum (dark/blue) of the 10th and maximum (light/yellow) of the 90th percentile over all values, time points and subjects. Top is for instantaneous PLI and bottom for GVD connectivity, averaged over subjects.}
	\label{face}
\end{figure}

Fig. \ref{face} shows the mean adjacency matrix over subjects ((a) top right) and the resulting $C_{loc}$ for instantaneous phase and GVD connectivity estimates, averaged over subjects. In the GVD connectivity, we can clearly see a strong pattern of dynamic connectivity in nodes 40-60 occurring around 0.3-0.5s after stimulus which dies away and then appears to return again near the 1s mark. This activity occurs after the N175 event-related potential known to play an important role in face perception tasks \cite{Rossion2011}, suggesting a post N175 phase-based functional response to the visual stimuli. Topoplots confirm that this is more evident using the GVD approach, Fig. \ref{face} (b), where a strong polarity of activity from front right to back left from 0.3-0.5s reoccurring at 0.9-1s is contrasted with a drop in activity from top left to back right. Activity from 0.3-0.5s is suggested also in the top left of instantaneous phase only but is less apparent.
	
\section{Conclusion}
We defined and provided a general framework for a new branch of multivariate signal analysis using graphs, termed graph-variate signal analysis. This concerned graph weighted instantaneous bivariate functions. We then elaborated on novel methodologies of graph-variate signal analysis towards the temporal-topological analysis of multivariate signals and reliable connectivity estimation at the resolution of the signal. In simulations we showed the robustness of the approach for finding correlations and detecting true activity within large datasets. In the latter it was shown to outperform similar state-of-the-art approaches. Pertinently, GVD connectivity excelled at differentiating coupling changes between EEG eyes-open and eyes-closed resting states and elucidating instantaneous phase-based activity in a face presentation task, compared to competitive approaches. These methods also showed promise in the interpretation and discovery of a wide range of datasets, including in geophysical well logs where our techniques could quickly identify areas and epochs of interest. We showed the unique setting occupied by this new form of analysis within a unified framework of multivariate signals and graphs. We also showed the limitations of using matrix multiplication of a graph adjacency matrix and graph signal vector, such as employed in GSP, for its formulation. We hope the methods and insights offered by this theory will be of use for numerous applications in analysing temporal dynamics of multivariate signals.
	
\section{Acknowledgements}
We would like to thank Phung T.K. Nguyen of the Edinburgh Time-Lapse Project, Heriot-Watt University, for her help in analysing and interpreting the geophysical data from the Kansas Geological Survey. We also thank the editor and reviewers for their generous feedback which amounted to a considerable contribution to this paper. 

\section*{Appendix: General unified framework of multivariate signals and graphs}	
To understand the broader context and relationships between analyses such as graph-variate signal analysis and GSP, it is required to construct a general, unified framework of multivariate signals and graphs. This framework can be implemented by studying the graph-variate signal object in Definition 1. Note that $\Gamma$ includes a multivariate signal associated with the node set similarly to how graph definitions usually include the weighted adjacency matrix associated with the edge set. Then
	\begin{itemize}
		\item $(\mathcal{V}, \mathbf{X})$ is the node space composed of a matrix $\mathbf{X}$ whose first dimension is indexed by the node set $\mathcal{V}$ and second dimension is indexed by a sequential characteristic of activity at the nodes, typically time.
		\item	$(\mathcal{E}, \mathbf{W})$ is the edge space composed of a weighted matrix $\mathbf{W}$ indexed by the edge set $\mathcal{E}$.
		\item $\Gamma$ constitutes the graph space of the combined node and edge spaces where nodes and edges joining those nodes are determined by the node labels $\{1,\dots, n\}$.
	\end{itemize}
The node space, being that which contains the activity at the nodes, frames the standard analysis of multivariate signals. Indeed, this is formalised by a general node function, $F_{\mathcal{V}}$, defined on the node space as
\begin{equation}\label{eq4}
	\begin{array}{cccc}
	F_{\mathcal{V}}: &\mathbb{R}^{n\times p} &\rightarrow &\mathbb{R}^{m\times q} \\
	&\mathbf{X} &\mapsto & F_{\mathcal{V}}(\mathbf{X}).
	\end{array}
\end{equation}
Useful examples of such functions where $n = m$ and $p=q$ include weight thresholds and spectral filtering functions, e.g. for bandpassing the signal in a frequency band of interest. Also, note that this terminology is adopted in the main body of the text in reference to the bivariate node function utilised in graph-variate signal analysis.
		
The edge space, on the other hand, is a topological space whose elements are the unlabelled isomorphism classes of graphs of size $n$ \cite{Bondy2008}. This is where one finds the standard analysis of networks. A function $F_{\mathcal{E}}$ on the edge space $(\mathcal{E},\mathbf{W})$ is defined on $\mathbb{R}^{n\times n}$ as 
\begin{equation}
	\begin{array}{cccc}
	F_{\mathcal{E}}: &\mathbb{R}^{n\times n} &\rightarrow &\mathbb{R}^{m\times l} \\
	&\mathbf{W} &\mapsto & F_{\mathcal{E}}(\mathbf{W}).
	\end{array}
\end{equation}
Some examples of such functions are
\begin{itemize}
	\item thresholds when $n=m=l$, 
	\item global network indices such as the clustering coefficient or characteristic path length, when $m = l = 1$,
	\item local network indices such as the local clustering coefficient or betweenness centrality, when $m = n$ and $l = 1$. 
\end{itemize}
These are necessarily all invariants under graph isomorphisms-- individuality of nodes is not considered.

Mappings from a space into the same space (e.g. from $\mathbb{R}$ to itself) are ubiquitous in mathematics and engineering and this is no different here. Indeed, we will see that such functions on the edge and node spaces take a prominent role in analyses of multivariate signals and graphs because they provide useful operations for acting on the reciprocal node and edge spaces. This is due to the matching inner dimensions of the adjacency matrix and the multivariate signal. Therefore, the following definitions will be useful.
\begin{definition}\label{preserve}
	An \textit{edge dimension preserving function}, $\bar F_{\mathcal{E}}$, maps the adjacency matrix, $\mathbf{W}\in \mathbb{R}^{n\times n}$, to a new matrix $\tilde{\mathbf{W}}\in\mathbb{R}^{n\times n}$.
\end{definition}
\begin{definition}
	A \textit{node dimension preserving function}, $\bar F_{\mathcal{V}}$, maps the multivariate signal, $\mathbf{X}\in \mathbb{R}^{n\times p}$, to a new signal $\tilde{\mathbf{X}}\in\mathbb{R}^{n\times p}$.
\end{definition}

We shall now consider how node and edge spaces can be combined to produce meaningful analyses for the graph-based analysis of multivariate signals. Particularly, in doing this we will comment on where GSP and graph-variate signal analysis methodologies occur.
	
\subsection{Edge-dependent operations acting on the node space}\label{edgeops}
Since the inner-dimensions of the edge space and node space agree, the output of any edge-dimension preserving function together with the usual matrix multiplication, $\cdot$, provide useful operations which act on the node space, $(\mathcal{V},\hat{\mathbf{X}})$:
\begin{equation}\label{Fe}
	\begin{array}{cccc}
	\bar{F}_{\mathcal{E}}(\mathbf{W})\cdot: &\mathbb{R}^{n\times p} &\rightarrow &\mathbb{R}^{n\times p} \\
	&\mathbf{X} &\mapsto & \bar{F}_{\mathcal{E}}(\mathbf{W})\cdot\mathbf{X}.
	\end{array}
\end{equation}
We thus realise that $\bar{F}_{\mathcal{E}}(\mathbf{W})\cdot$ is in fact a node dimension preserving function. Some of the simplest examples include the weighted adjacency matrix, $\mathbf{W}$, and the graph Laplacian, $\mathbf{L}$. Note, this is precisely where the various aspects of GSP can be framed in the general framework. This can be seen since important definitions in GSP involve pre-matrix multiplication of the graph signal by matrices derived from graphs. For example, the GFT treats the eigenvectors of the Laplacian or the graph adjacency matrix as a basis for the decomposition of graph signals into graph frequency components. The $l$th eigenvector produces the $l$\textsuperscript{th} frequency component of the graph signal, $\mathbf{x}\in\mathbb{R}^{n\times 1}$, defined as $\mathbf{u}_{l}\cdot\mathbf{x}$. Similarly, graph convolution, translation, modulation and graph wavelets can be formulated as matrix multiplication on linear components of the graph signal. Further, polynomials of the adjacency and Laplacian matrices are implemented to construct graph signal filters in GSP in \cite{Sand2013} and \cite{Shum2013}, respectively, which are then matrix multiplied by the graph signal.
	
\subsection{Node-dependent operations acting on the edge space}
Because the edge space is composed of pairs of elements in the node space, when combining the output of node space functions with the adjacency matrix it is most sensible to impose that the elements acting on the weight $w_{ij}$ be bivariate functions of the signal at nodes $i$ and $j$. Using this function in a dimension preserving mapping from the edge space to itself, reciprocating \eqref{Fe} for node spaces, we can use the Hadamard product on the tensor $\underline{\mathbf{J}} = F_{\mathcal{V}}(\mathbf{X})$, as described in Section II.A. This gives
\begin{equation}
	\begin{array}{cccc}
	\circ F_{\mathcal{V}}(\mathbf{X}): &\mathbb{R}^{n\times n} &\rightarrow &\mathbb{R}^{n\times n \times p} \\
	&\mathbf{W} &\mapsto & \mathbf{W}\circ F_{\mathcal{V}}(\mathbf{X}).
	\end{array}
\end{equation}

In fact, it is exactly in this manner that we define graph-variate signal analysis in Definition 2. Thus, we have established that GSP and graph-variate signal analysis appear to be framed in different places within the general framework outlined. Still, both edge-dependent operations acting on the node space and node-dependent operations acting on the edge space clearly end up with outputs which contain combinations of node space and edge space elements. Therefore, the overlap between these two kinds of analysis remains to be seen. For instance, perhaps framing particularly complicated edge space functions $\bar{F}_{\mathcal{E}}$ through some clever combination of graph Laplacians may provide the full array of graph-variate signal analyses. In which case, the distinction between graph-variate signal analysis and GSP would be purely conceptual.

We shall thus prove that irreconcilable differences exist between the analyses outlined in section A and section B of this Appendix. For this, we require to pose an output of graph-variate signal analysis which has the same dimensions and relates to the same components as the output of \eqref{Fe}. This requires us to define a new operator which allows node space operations to act on the edge space to provide local graph-variate analysis for each node as follows.
\begin{definition}
For a matrix $\mathbf{A}\in\mathbb{R}^{n\times n}$ and 3D tensor $\underline{\mathbf{B}}\in\mathbb{R}^{n\times n\times p}$, composed of the $p$ $n\times n$ matrices $\{\underline{\mathbf{B}}_{(t)}\}_{t = 1}^{p}$, their \textbf{signal product}, $\mathbf{A} \diamond \underline{\mathbf{B}}$, is the matrix whose $t$\textsuperscript{th} column is the vector $[\sum_{j}A_{ij}B_{jit}]_{i= 1}^{n}$, which is the dot product of the $i$th rows of $\mathbf{A}$ and the $i$th columns of $\underline{\mathbf{B}}_{(t)}$.
\end{definition}
Then
\begin{equation}\label{diamond}
(\mathbf{W}\diamond\underline{\mathbf{J}})_{it} = \sum_{j=1}^{n}w_{ij}F_{\mathcal{V}}(x_{i}(t),x_{j}(t)).
\end{equation}	
This is equivalent to taking the sum of the rows of $\mathbf{W}\circ\underline{\mathbf{J}}$, using the mode-$k$ Hadamard product. A special case of this is GSP's node gradient formula \cite{Shum2013} where $F_{\mathcal{V}}(x_{i}(t),x_{j}(t)) = (x_{i}(t)-x_{j}(t))^{2}$. 

It is straightforward to note that node space functions $x_{j}(t)$ and $x_{i}(t)-x_{j}(t)$ are solutions for $F_{\mathcal{V}}$ in \eqref{diamond} to the equations $\mathbf{W}\diamond\underline{\mathbf{J}} = \mathbf{W}\cdot\mathbf{X}$ and $\mathbf{W}\diamond\underline{\mathbf{J}} = \mathbf{L}\cdot\mathbf{X}$, respectively. But what can be said of general operators $\bar{F}_{\mathcal{E}}$? This is answered in the following proposition.
\begin{propapp}
	For the output of an edge dimension preserving function $\bar{F}_{\mathcal{E}}(\mathbf{W})$ and of a node function $F_{\mathcal{V}}(\mathbf{X})$,
	\begin{equation}
	\bar{F}_{\mathcal{E}}(\mathbf{W})\cdot\mathbf{X} = \mathbf{W}\diamond F_{\mathcal{V}}(\mathbf{X})
	\end{equation}
	if and only if $F_{\mathcal{V}}(\mathbf{X}) = a_{ij}x_{i}(t) + a_{ji}x_{j}(t)$ for some constants $a_{ij},a_{ji}\in\mathbb{R}$, and
	\begin{equation}
	\bar{F}_{\mathcal{E}}(\mathbf{W}) = 
	\begin{bmatrix}
	\sum_{j}a_{1j}w_{1j} & a_{21}w_{12} & \dots  & a_{n1}w_{1n} \\
	a_{12}w_{21} & \sum_{j}a_{2j}w_{2j} & \dots  & a_{n2}w_{2n} \\
	\vdots & \vdots & \ddots & \vdots \\
	a_{1n}w_{n1} & a_{2n}w_{n2} & \dots  & \sum_{j}a_{nj}w_{nj} 
	\end{bmatrix}.
	\end{equation}
	\begin{proof}
		We first note that matrix multiplication with $\mathbf{X}$ is linear on the entries of $\mathbf{X}$ thus we cannot consider equating $\bar{F}_{\mathcal{E}}(\mathbf{W})\cdot\mathbf{X}$ to a graph weighted non-linear node space function-- one cannot obtain elements $x_{i}(t)^{p}$ for $p>1$. Further, since each element of $\bar{F}_{\mathcal{E}}(\mathbf{W})$ is multiplied by an element of $\mathbf{X}$ and each element of $\bar{F}_{\mathcal{V}}(\mathbf{X})$ is multiplied by an entry of $\mathbf{W}$, there can be no constants in either function.
		
		Now, in the linear case without constants for $\mathbf{x}\in\mathbb{R}^{n\times 1}$,
		\begin{equation*}
		(\widetilde{\mathbf{W}}\cdot\mathbf{X})_{ti} = \sum_{j=1}^{n}w_{ij}(a_{ij}x_{i}(t)+ a_{ji}x_{j}(t))\\
		\end{equation*}
		\begin{equation}
		\iff \widetilde{w}_{ij} = \left\{
		\begin{array}{cc}
		\sum_{p=1}^{n}a_{ip}w_{ip} & i = j \\
		a_{ji}w_{ij} & i\neq j,\\
		\end{array}
		\right.
		\end{equation}
		for coefficients $a_{ij}\in\mathbb{R}$, satisfying the proposition.
	\end{proof}
\end{propapp}
This result then proves the methodological novelty of graph-variate signal analysis.

Of course, one obvious remaining aspect to consider in this framework is the combination of GSP and graph-variate signal analysis approaches. For this we can consider methods posed by the following formula:
\begin{equation}
\bar{F}_{\mathcal{E}}(\mathbf{W})\circ F_{\mathcal{V}}(\mathbf{X}),
\end{equation}
It lies out of the scope of this work to look into this, but it remains as a potentially fruitful avenue for future research.

\end{document}